\documentclass[aps,prl,showpacs,twocolumn,superscriptaddress]{revtex4}
\usepackage{amsmath}
\usepackage{amssymb}
\usepackage{epsfig}
\usepackage{graphicx,amsmath}

\begin{document}

\title{Quantum critical point in high-temperature superconductors}
\author{V.R. Shaginyan}\email{vrshag@thd.pnpi.spb.ru}
\affiliation{Petersburg Nuclear Physics Institute, RAS, Gatchina,
188300, Russia}\affiliation{Racah Institute of Physics, Hebrew
University, Jerusalem 91904, Israel}
\author{M.Ya. Amusia}\affiliation{Racah Institute
of Physics, Hebrew University, Jerusalem 91904, Israel}
\author{K.G. Popov}\affiliation{Komi Science Center, Ural Division,
RAS, Syktyvkar, 167982, Russia}
\author{V.A. Stephanovich}\affiliation{Opole University, Institute of Mathematics
and Informatics, Opole, 45-052,Poland}\email{stef@math.uni.opole.pl}

\begin{abstract}
Recently, in high-$T_c$ superconductors (HTSC), exciting
measurements have been performed  revealing their physics in
superconducting and pseudogap states and in normal one induced by
the application of magnetic field, when the transition from
non-Fermi liquid to Landau Fermi liquid behavior occurs. We employ
a theory, based on fermion condensation quantum phase transition
which is able to explain facts obtained in the measurements. We
also show, that in spite of very different microscopic nature of
HTSC, heavy-fermion metals and 2D $\rm ^3He$, the physical
properties of these three classes of substances are similar to each
other.
\end{abstract}

 \maketitle

\section{Introduction}

Non-Fermi liquid (NFL) behavior of many classes of strongly
correlated fermion systems projects one of the tremendous
challenges in modern condensed matter physics. This behavior is so
unusual that the traditional Landau paradigm of quasiparticles does
not apply to it. It is widely believed that utterly new concepts
are required to describe the underlying physics
\cite{senth,senth1,col,lohneysen,si,sach}. There is a fundamental
question: how many concepts do we need to describe the above
physical mechanisms? This cannot be answered on purely experimental
or theoretical grounds. Rather, we have to use both of them.
Recently, in high-$T_c$ superconductors (HTSC), exciting
measurements revealing their physics have been performed. One type
of the measurements demonstrate the existence of Bogoliubov
quasiparticles (BQ) in a superconducting state
\cite{mat,norm,vall}. At the same time, in the pseudogap regime at
$T > T_c$ (when the superconductivity vanishes), a strong
indication of the pairing of electrons or the formation of
preformed electron pairs has been observed, while the gap continues
to follow a simple d-wave form \cite{norm,vall}. Another type of
the measurements explore the normal state induced by the
application of magnetic field, when transition from NFL to Landau
Fermi liquid (LFL) behavior occurs \cite{pnas}. This transition
takes place at magnetic field $B\geq B_{c2}\geq B_{c0}$, where
$B_{c2}$ is the field destroying the superconducting state, and
$B_{c0}$ is the critical field at which the magnetic field induced
quantum critical point (QCP) takes place \cite{pnas}. We note that
to study the aforementioned transition experimentally, the strong
magnetic fields $B\geq B_{c2}$ are required so that earlier such
investigation was technically inaccessible. An attempt to study the
transition experimentally had been done more than 10 years ago
\cite{mack}.

Many puzzling and common experimental features of such seemingly
different systems as two-dimensional (2D) electron systems and
liquid $^3$He, heavy-fermion (HF) metals and HTSC suggest that
there is a hidden phase transition, which remains to be recognized.
The key word here is quantum criticality, taking place at QCP.
Heavy fermion metals provide important examples of strongly
correlated Fermi-systems. The second class of substances to test
whether or not Landau quasiparticles \cite{lan} play an underlying
role to construct the superconducting state and to form BQ on their
base in accordance with Bardeen-Cooper-Schrieffer (BCS) theory
\cite{bcs}, are HTSC. In these substances, all QCP are almost
inaccessible to experimental observations since they are "covered"
by superconductivity. More precisely, the superconductive gap
opened at the Fermi level, changes the physical properties of
corresponding quantum phase transition.

There is a common wisdom that the physical properties of above
systems are related to zero temperature quantum fluctuations,
suppressing quasiparticles and thus generating their NFL properties
\cite{senth,senth1,col,lohneysen,si,sach}, depending on their
ground state, either magnetic or superconductive. On the other
hand, it was shown that the electronic system of HF metals
demonstrates the universal low-temperature behavior irrespectively
of their magnetic ground state \cite{epl2}. Recently, the NFL
behavior has been discovered experimentally in 2D $^3$He
\cite{he3}, and the theoretical explanation has been given to it
\cite{prlmy}, revealing the similarity in physical properties of 2D
$^3$He and HF metals. We note here that $^3$He consists of neutral
atoms interacting via van der Waals forces, while the mass of He
atom is 3 orders of magnitude larger than that of an electron,
making $\rm ^3He$ to have drastically different microscopic
properties than those of HF metals. Therefore it is of crucial
importance to check whether this behavior can be observed in other
Fermi systems like HTSC. As we shall see, the precise measurements
on HTSC's Bi$_2$Sr$_2$Ca$_2$Cu$_3$O$_{10+x}$ \cite{mat}, $\rm
Bi_2Sr_2CaCu_2O_{8+x}$ \cite{vall} and Tl$_2$Ba$_2$CuO$_{6+x}$
\cite{pnas} allow us to establish the relationships between
physical properties of both HTSC compounds and HF metals and
clarify the role of Landau quasiparticles.

In this letter, we consider a superconducting state of HTSC in the
framework of our theory based on the fermion condensation quantum
phase transition (FCQPT) concept.  We show that the superconducting
state is BCS-like, the elementary excitations are BQ, and the
primary ideas of the LFL  and BCS  theories remain valid, whereas
the maximal value of a superconducting gap and other exotic
properties are determined by the presence of underlying fermion
condensate (FC).  This presence manifests itself in a fact that the
quasiparticle effective mass $M^*$ strongly depends on temperature,
magnetic field and doping $x$. We show, that in spite of very
different microscopic nature of HTSC, HF metals and 2D $\rm ^3 He$,
their physical properties belong to universal behavior of strongly
correlated Fermi-systems. We demonstrate that the physics
underlying the field-induced reentrance of LFL behavior is the same
for HTSC compounds and HF metals. We demonstrate that there is at
least one quantum phase transition inside the superconducting dome,
and this transition is indeed FCQPT. We also show that there is a
relationship between the critical fields $B_{c2}$ and  $B_{c0}$ so
that $B_{c2}\gtrsim B_{c0}$.

\section{Superconducting and pseudogap states}

At $T<T_c$, the thermodynamic potential $\Omega$ of an electron
liquid is given the equation (see, e.g. \cite{til})
\begin{equation}\label{1}
\Omega = E_{\rm gs}-\mu N-TS,
\end{equation} where $N$ is particles number, $S$ denotes
the entropy, and $\mu$ is a chemical potential. The ground state
energy $E_{\rm gs}[\kappa({\bf p}),n({\bf p})]$ of an electron
liquid is a functional of superconducting order parameter
$\kappa({\bf p})$ and of the quasiparticle occupation numbers
$n({\bf p})$. Here we assume that the electron system is
two-dimensional, while all results can be easily generalized to the
case of three-dimensional system. The energy $E_{\rm gs}$ is
determined by the standard equation of the weak-coupling theory of
superconductivity
\begin{equation}\label{2}
 E_{\rm gs}\ =\ E[n({\bf p})]+\int \lambda_0V({\bf
p}_1,{\bf p}_2) \kappa({\bf p}_1) \kappa^*({\bf p}_2) \frac{d{\bf
p}_1d{\bf p}_2}{(2\pi)^4}.
\end{equation}
Here $E[n({\bf p})]$ is the Landau functional determining the
ground-state energy of a normal Fermi liquid. Here $\lambda_0V$ is
the pairing interaction and $\lambda_0$ is the coupling constant.
Here
\begin{equation}\label{3}
 n({\bf p})=v^2({\bf p})\left[1-f({\bf p})\right]+u^2({\bf
p})f({\bf p}) ,
\end{equation} and
\begin{equation}\label{4}
\kappa({\bf p})=v({\bf p})u({\bf p})\left[1-2f({\bf p})\right],
\end{equation} where
the coherence factors $v({\bf p})$ and $u({\bf p})$ are obeyed the
normalization condition \begin{equation}\label{5}
 v^2({\bf p})+u^2({\bf
p})=1.
\end{equation}
The distribution function $f({\bf p})$ of BQ
defines the entropy
\begin{equation}\label{6}
S=-2\int\left[f({\bf
p})\ln f({\bf p})+(1-f({\bf p}))\ln(1-f({\bf p}))\right]\frac{d{\bf
p}}{4\pi^2}.
\end{equation}
We assume that the pairing interaction $\lambda_0V$ is weak and
produced, for instance, by electron-phonon interaction. Minimizing
$\Omega$ with respect to $\kappa({\bf p})$ and using the definition
$\Delta({\bf p})=-\delta \Omega/\kappa({\bf p})$, we obtain
\begin{equation}\label{7a}
\Delta({\bf p})=-\int\lambda_0V({\bf p},{\bf p}_1)\kappa({\bf p}_1)
\frac{d{\bf p}_1}{(2\pi)^2},\end{equation}
\begin{equation}\label{7} \varepsilon({\bf p})-\mu=\Delta({\bf p})\frac{1-2v^2({\bf
p})} {2v({\bf p})u({\bf p})}.
\end{equation} The single-particle
energy $\varepsilon({\bf p})$ is determined by the Landau equation
\begin{equation}\label{8}
 \varepsilon({\bf p})= \frac{\delta E[n({\bf
p})]}{\delta n({\bf p})}.
\end{equation}
Note that $E[n({\bf p})]$, $\varepsilon[n({\bf p})]$, and the Landau
amplitude
\begin{equation}\label{9}
F_L({\bf p},{\bf p}_1)=\frac{\delta^2 E[n({\bf p})]}{\delta n({\bf
p})\delta n({\bf p}_1)}
\end{equation}
depend implicitly on the density $x$ which defines the strength of
$F_L$. Minimizing $\Omega$ with respect to $f({\bf p})$ and after
some algebra, we obtain the following explicit equation for the
superconducting gap $\Delta({\bf p})$
\begin{equation}\label{10}
 \Delta({\bf p})= -\frac{1}{2}\int\lambda_0 V({\bf
p},{\bf p}_1) \frac{\Delta({\bf p}_1)}{E({\bf p}_1)}\left[1-2f({\bf
p}_1)\right]) \frac{d{\bf p}_1}{4\pi^2}.
\end{equation} Here the excitation
energy $E({\bf p})$ of BQ is given by
\begin{equation}\label{11}
 E({\bf p})=\frac{\delta (E_{gs}-\mu N)}{\delta f({\bf p})}=
\sqrt{(\varepsilon({\bf p})-\mu)^2+\Delta^2({\bf p})}.
\end{equation} The coherence factors  $v({\bf p})$, $u({\bf p})$,
and the distribution function $f({\bf p})$ are given by the ordinary
relations
\begin{eqnarray}\label{12}
&&v^2({\bf p})=\frac{1}{2}\left[1-\xi({\bf p})\right],\ \ u^2({\bf
p})=\frac{1}{2}\left[1+\xi({\bf p})\right], \\
&&\xi({\bf p})=\frac{\varepsilon({\bf p})-\mu}{E({\bf p})},\ \
 f({\bf p})=\frac{1}{1+\exp(E({\bf p})/T)}. \label{13}
\end{eqnarray}

The equations (\ref{7})-(\ref{13}) are conventional BSC equations
\cite{bcs,til} for the superconducting state with BQ and the maximal
value of the superconducting gap $\Delta_1\propto
\exp(-1/\lambda_0)$ if the system in question has not undergone
FCQPT.

Now we consider a superconducting electron liquid with FC taking
place after FCQPT point. If $T=0$ and $\lambda_0\to 0$, then both
maximal value of the superconducting gap $\Delta_1\to 0$ and the
critical temperature $T_c\to 0$ so that Eq. \eqref{7} reduces to the
equation \cite{xoshag,obz1,volovik,khod,amshag,obz,khodb}
\begin{equation}\label{FCM}
\frac{\delta E}{\delta n({\bf
p})}=\varepsilon({\bf p})=\mu,\, {\rm if}\, 0\leq n({\bf p})\leq1;\:
p_i\leq p\leq p_f\
\end{equation}
provided that the order parameter $\kappa$ is finite at $p_i\leq
p\leq p_f$. Equation \eqref{FCM} defines a new state of electron
liquid with FC characterized by a flat part of the spectrum in the
$p_f-p_i$ region. This state has a strong impact on the system
properties and emerges at some critical density $x=x_{FC}$ where
the amplitude $F_L$ becomes strong enough. On the contrary, when
the Landau amplitude $F_L(p=p_F,p_1=p_F)$ as a function of density
$x$ is sufficiently small, the flat part vanishes, and at $T\to 0$
Eq. \eqref{FCM} has the only trivial solution
$\varepsilon(p=p_F)=\mu$ so that the quasiparticle occupation
numbers are given by a step function, $n({\bf p})= \theta(p_F-p)$.

\begin{figure} [! ht]
\begin{center}
%\vspace*{-2cm}
\includegraphics [width=0.45\textwidth]{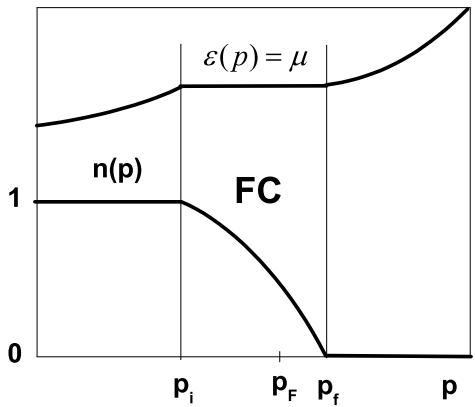}
\end{center}
\caption{Schematic plot of the quasiparticle occupation number
$n(p)$ and spectrum $\varepsilon(p)$ in the FC state. Function
$n(p)$ obeys the relations $n(p \leq p_i) = 1$, $n(p_i < p < p_f) <
1$ and $n(p \geq p_f) = 0$, while $\varepsilon (p_i < p < p_f) =
\mu$. Fermi momentum $p_F$ satisfies the condition $p_i < p_F <
p_f$.}\label{ff1}
\end{figure}

Upon applying well-known Landau equation, we can relate a
quasiparticle effective mass $M^*$ to the bare electron mass $M$
\cite{lan,pfit}
\begin{equation}\label{MM*}
\frac{M^*}{M}=\frac{1}{1-N_0F^1(x)/3}.
\end{equation}
Here $N_0$ is the density of states of a free electron gas, $x
=p_F^3/3\pi^2$ is a number density, $p_F$ is Fermi momentum, and
$F^1(x)$ is the $p$-wave component of Landau interaction $F_L$. When
at critical density $x=x_{FC}$, $F^1(x)$ achieves a threshold value,
the denominator in Eq. \eqref{MM*} tends to zero so that the
effective mass diverges at $T=0$ and the system undergoes FCQPT. The
leading term of this divergence reads
\begin{equation}\label{M1M}
\frac{M^*(x)}{M}=\alpha_1+\frac{\alpha_2}{x-x_{FC}},
\end{equation}
where $\alpha_1$ and $\alpha_2$ are constants. At $x<x_{FC}$ the FC
takes place. The essence of this phenomenon is that at $x<x_c$ the
effective mass \eqref{M1M} becomes negative signifying physically
meaningless state. To avoid this state, the system reconstructs its
quasiparticle occupation number $n({\bf p})$ and topological
structure so as to minimize its ground state energy $E$. The main
result of such reconstruction is that instead of Fermi step, we have
$0\leq n(p)\leq 1$ in certain range of momenta $p_i\leq p\leq p_f$,
see Eq. \eqref{FCM}. Accordingly, in the above momenta interval, the
spectrum $\varepsilon (p)=\mu$, see Fig. \ref{ff1} for details of
its modification.

Due to above peculiarities of the $n({\bf p})$ function, FC state is
characterized by the superconducting order parameter $\kappa({\bf
p})=\sqrt{n({\bf p})(1-n({\bf p}))}$. This means that if the
electron system with FC has pairing interaction with coupling
constant $\lambda$, it exhibits superconductivity since as it
follows from Eq. \eqref{7a} $\Delta_1\propto\lambda$ in a weak
coupling limit. This linear dependence is also a peculiarity of FC
state \cite{xoshag,obz1,obz,amshag} and substitutes above well-known
BCS relation $\Delta _1\propto\exp{(-1/\lambda _0)}$.

Now we can study the relation between the state defined by Eq.
\eqref{FCM} and the superconductivity. At $T\to0$, Eq. \eqref{FCM}
defines a particular state of a Fermi liquid with FC, for which the
modulus of the order parameter $|\kappa({\bf p})|$ has finite values
in the $(p_f-p_i)$ region, whereas  $\Delta_1\to 0$ in this region.
We observe that $f({\bf p},T\to0)\to0$, and it follows from Eqs.
\eqref{3} and \eqref{4} that $0<n({\bf p})<1$ implies that
$|\kappa({\bf p})|\neq 0$ in the region $(p_f-p_i)$ . Such a state
can be considered as superconducting with an infinitely small value
of $\Delta_1$ so that the entropy of this state is equal to zero. It
is obvious that this state being driven by the quantum phase
transition disappears at $T>0$ \cite{amshag}. Any quantum phase
transition at $T=0$ is determined by a control parameter other then
temperature, for example, by pressure, by magnetic field, or by the
density $x$ of mobile charge carriers. Since a quantum phase
transition occurs at a QCP, in our FCQPT case at $T=0$ the role of
QCP is played by critical density $x = x_{FC}$.

If $\lambda_0\neq 0$, then $\Delta_1$ becomes finite. It is seen
from Eq. \eqref{10} that the superconducting gap depends on the
single-particle spectrum $\varepsilon({\bf p})$. On the other hand,
it follows from Eq. \eqref{7} that $\varepsilon({\bf p})$ depends
on $\Delta({\bf p})$ provided that at $\Delta_1\to 0$ Eq.
\eqref{FCM} has the solution corresponding to FC existence. Let us
assume that $\lambda_0$ is small so that the BSC interaction
$\lambda_0 V({\bf p},{\bf p}_1)$ can only lead to a small
correction to the order parameter $\kappa({\bf p})$ determined by
Eq. \eqref{FCM}. Upon differentiation both parts of Eq. \eqref{7}
over momentum $p$, we obtain that $M^*$ becomes finite
\begin{equation}\label{17}
 M^*\sim p_F\frac{p_f-p_i}{2\Delta_1}.
\end{equation}
It follows from Eq. \eqref{17} that the effective mass and the
density of states $N(0)\propto M^*\propto 1/\Delta_1$ are finite
and constant at $T<T_c$. As a result, we conclude that in contrast
to the conventional theory of superconductivity the single-particle
spectrum $\varepsilon({\bf p})$ strongly depends on the
superconducting gap and we have to solve Eqs. \eqref{8} and
\eqref{10} self-consistently. On the other hand, let us assume that
Eqs. \eqref{8} and \eqref{10} are solved, and the effective mass
$M^*$ is determined. Now one can fix the dispersion
$\varepsilon({\bf p})$ by choosing the effective mass $M^*$ of
system in question equal to $M^*_{FC}$ and then solve Eq.
\eqref{10} as it is done in the case of the conventional theory of
superconductivity \cite{bcs}. As a result, one observes that the
superconducting state is characterized by BQ with the dispersion
\eqref{11}, the coherence factors  $u$, $v$ \eqref{12} and
normalization condition \eqref{5}. Thus, the observed features
agree with BQ behavior predicted by BCS theory. This suggests that
at $T\leq T_c$ the superconducting state with FC is BCS-like and
implies the basic validity of BCS formalism for the description of
this state. It is exactly the case observed experimentally in
HTSC's Bi$_2$Sr$_2$Ca$_2$Cu$_3$O$_{10+x}$ and $\rm
Bi_2Sr_2CaCu_2O_{8+x}$ \cite{mat,vall}.

\begin{figure} [! ht]
\begin{center}
\includegraphics [width=0.47\textwidth] {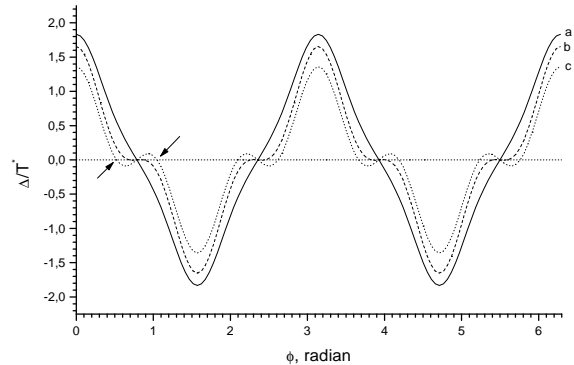}
\end{center}
\caption {The gap $\Delta(p_F,\phi)$ as a function of $\phi$
calculated for three values of the temperature $T$ in units of
$T_{\rm node} \simeq T_c$. The curve (a) (solid line) represents the
calculation for $T=0.9T_{\rm node}$, the curve (b) (dashed line)
represents the same at $T=T_{\rm node}$ and the curve (c) (dotted
line) reports the calculation at $T=1.2T_{\rm node}$. The arrows
indicate the region $\theta_c$ limited by the two new zeros emerging
at $T>T_{\rm node}$.} \label{Fig2}
\end{figure}

It has been shown in Refs \cite{obz,ars}, that in the presence of
FC, Eq. \eqref{10} has nontrivial solutions at $T<T^*$ ($T^*$ is a
temperature at which Eq. \eqref{10} has only trivial solution
$\Delta=0$) when the pairing interaction $\lambda_0V$ consists of
attraction and strong repulsion leading to $d$-wave
superconductivity. At some temperature $T_{\rm node}$, the gap
$\Delta({\bf p})$ as a function of the angle $\phi$ ($\Delta({\bf
p})=\Delta(p_F,\phi)$) obtains new nodes as shown in Fig. \ref{Fig2}
\cite{ars}.
Figure \ref{Fig2} shows the ratio $\Delta (p_F,\phi)/T^*$ calculated
for three temperatures: $0.9\,T_{\rm node}$, $T_{\rm node}$ and $1.2
\,T_{\rm node}$. In contrast to curve (a), curves (b) and (c) have
approximately flat sections. Clearly, the flattening occurs due to
new zeros emerging at $T=T_{\rm node}$, see Fig. \ref{Fig2}. As the
temperature increases, the region $\theta_c$ between zeros
(indicated by arrows in Fig. \ref{Fig2}) increases in size. It is
also clear that the gap $\Delta$ is very small within the interval
$\theta_c$. Thus, we conclude that the gap in the vicinity of
$\theta_c$ can be destroyed at $T\geq T_c$ by any strong
fluctuations (e.g. antiferromagnetic), impurities, and sizable
inhomogeneities existing in HTSC. Since the superconducting gap is
destroyed in a macroscopic region of the phase space, $\theta_c$,
the coherence necessary for superconductivity is vanished and
superconductivity is also destroyed. This observation allows us to
conclude that $T_c\simeq T_{\rm node}$, while at $T\geq T_c$ the
pseudogap is formed. The behavior and the shape of the pseudogap
resembles closely similar characteristics of the superconducting gap
as Fig. \ref{Fig2} shows. The main difference is that the pseudogap
disappears in the segment $\theta_c$ of the Fermi surface, while the
gap disappears at isolated nodes of the $d$-wave. Our estimates show
that for small values of the angle $\psi$, the function
$\theta_c(\psi)$ rapidly increases,
$\theta_c(\psi)\simeq\sqrt{\psi}$. These estimates agree with the
results of numerical calculations of the function
$\theta_c([T-T_c]/T_c)$.

At temperatures $T<T_c $, the quasiparticle excitations of the
superconducting state are characterized by the presence of sharp
peaks. When the temperature becomes high ($T>T_c$) and
$\Delta(\theta)\equiv 0 $ in the interval $\theta_c$, normal
quasiparticle excitations with a width $\gamma$ appear in the
segments $\theta_c$ of the Fermi surface. A pseudogap exists outside
the segments $\theta_c$, and the Fermi surface is occupied by BQ in
this region. Excitations of both types have widths of the same order
of magnitude, transferring their energy and momenta into excitations
of normal quasiparticles. We estimate the value of $\gamma$. If the
entire Fermi surface were occupied by the normal state, the width
$\gamma$ would be $\gamma\approx N(0)^3T^2/\epsilon(T)^2$ with the
density of states $N(0)\sim M^*(T)\sim1/T$. The dielectric constant
$\epsilon(T)\sim N(0)$ and hence $\gamma\sim T$ \cite{obz1}.
However, only a part of the Fermi surface within $\theta_c$ is
occupied by normal excitations in our case. Therefore, the number of
states accessible for quasiparticles and quasiholes is proportional
to $\theta_c$, and the factor $T^2$ is replaced by the factor
$T^2\theta_c^2$. Taking all this into account yields $\gamma\sim
\theta_c^2 T\sim T(T-T_c)/T_c\sim(T-T_c)$. Here, we ignored the
small contribution from BCS - type excitations. It is precisely for
this reason that the width $\gamma$ vanishes at $T=T_c$, while the
resistivity of the normal state $\rho(T)\propto
\gamma\propto(T-T_c)$, because $\gamma\sim T-T_c$.

\section{General properties of heavy-fermion metals}

We have shown earlier (see, e.g. \cite{obz}) that without loss of
generality, to study the above universal behavior, it is sufficient
to use the simplest possible model of a homogeneous heavy-electron
(fermion) liquid. This permits not only to better reveal the
physical nature of observed effects, but to avoid unnecessary
complications related to microscopic features (like crystalline
structure, defects and impurities etc) of specific substances.

Now we consider the action of external magnetic field on HF liquid
in FC phase. Assume now that $\lambda_0$ is infinitely small. Any
infinitesimal magnetic field $B \neq 0$ (better to say, $B\geq
B_{c0}$) destroys both superconductivity and FC state, splitting it
by Landau levels. The simple qualitative arguments can be used to
guess what happens to FC state in this case. On one side, the energy
gain from FC state destruction is $\Delta E_B \propto B^2$ (see
above) and tends to zero as $B \to 0$. On the other side, $n(p)$ in
the interval $p_i\leq p\leq p_f$ gives a finite energy gain as
compared to the ground state energy of a normal Fermi liquid
\cite{obz}. It turns out that the state with largest possible energy
gain is formed by a multiconnected Fermi surface, so that the smooth
function $n(p)$ is replaced in the interval $p_i\leq p\leq p_f$ by
the set of rectangular blocks of unit height, formed from Heavyside
step functions \cite{khodb,zb,pog}. In this state the system
demonstrates LFL behavior, while the effective mass strongly depends
on magnetic field \cite{obz,pog},
\begin{equation}\label{MB}
M^*(B)\propto \frac{1}{\sqrt{B-B_{c0}}}.
\end{equation}
Here $B_{c0}$ is the critical magnetic field driving corresponding
QCP towards $T=0$. In some cases, for example in HF metal
CeRu$_2$Si$_2$, $B_{c0}=0$, see e.g. \cite{takah}. In our simple
model $B_{c0}$ is taken as a parameter.

\begin{figure}[!ht]
\begin{center}
\vspace*{-0.5cm}
\includegraphics [width=0.48\textwidth]{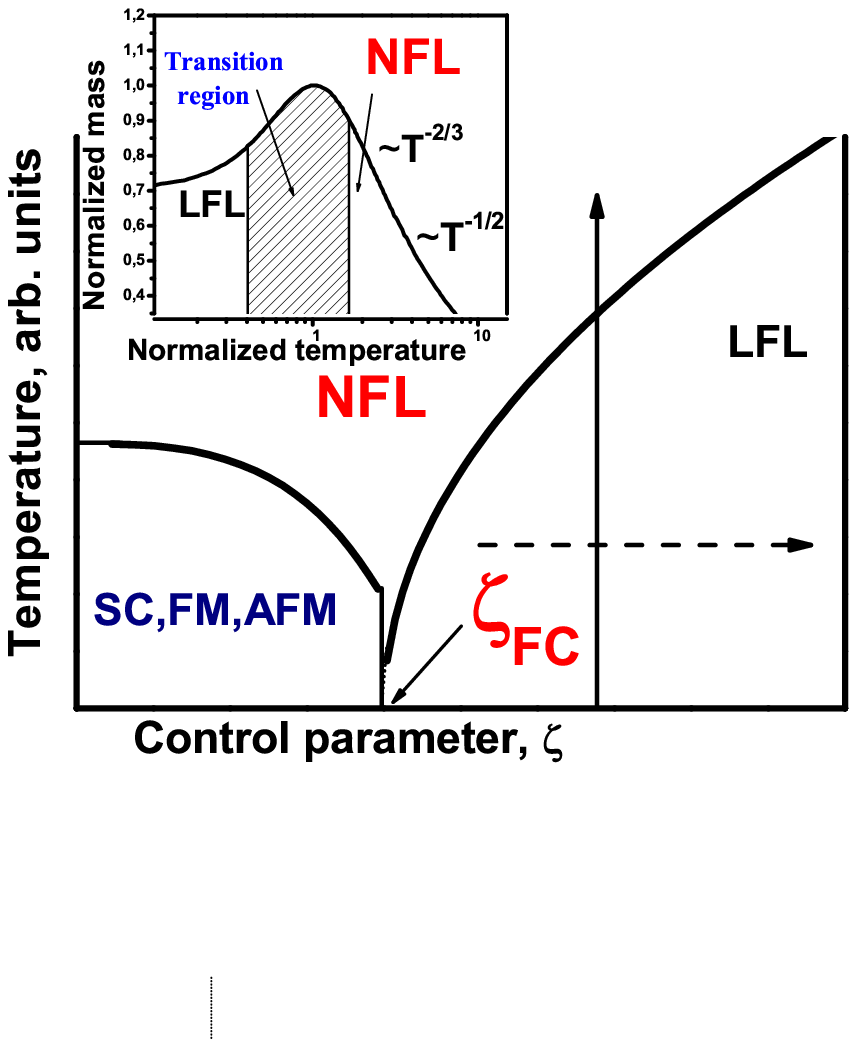}
\vspace*{-3.5cm}
\end{center}
\caption{Schematic phase diagram of HF metal. Control parameter
$\zeta$ represents doping $x$, magnetic field $B$, pressure $P$ etc.
$\zeta_{\rm FC}$ denotes the point of effective mass divergence. The
vertical arrow shows LFL-NFL transitions at fixed $\zeta$ with $M^*$
depending on $T_N$ as given by Eq. \eqref{fin1}. The dash horizontal
arrow illustrates the system moving in LFL regime along $\zeta$ at
fixed $T_N$, while $M^*(B)$ is given by Eq. \eqref{MB}. At $\zeta
<\zeta_{\rm FC}$ the system can be in a superconducting (SC),
ferromagnetic (FM) or antiferromagnetic (AFM) states. Inset shows a
schematic plot of the normalized effective mass versus the
normalized temperature. Transition regime, where $M^*_N$ reaches its
maximum, is shown by the hatched area. }\label{PHD}
\end{figure}

At elevated temperatures, the system transits from the LFL to NFL
regime as shown by the solid vertical arrow in Fig. \ref{PHD}
exhibiting the low-temperature universal behavior independent of its
magnetic ground state, composition, dimensionality (2D or 3D) and
even nature of constituent Fermi particles which may be electrons or
$\rm ^3He$ atoms \cite{epl2,prlmy}. To check, whether the
quasiparticles are present in the systems in the transition regime,
we use the results of measurements of heat capacity $C$, entropy $S$
and magnetic susceptibility $\chi$. If these results can be fitted
by the well-known relations from Fermi liquid theory
$C/T=\gamma_0\propto S/T\propto \chi\propto M^*$, then
quasiparticles define the system properties in the transition
regime.

Consider temperature and magnetic field dependence of the effective
mass $M^*(T,B)$ as system approaches FCQPT. Landau equation
\cite{lan} is of the form
\begin{equation}\label{sam2}
\frac{1}{M^*}=\frac{1}{M}+\int \frac{{\bf p}_F{\bf p}_1}{p_F^3}
F_L({\bf p_F},{\bf p}_1)\frac{\partial n(p_1,T,B)}{\partial p_1}
\frac{d{\bf p}_1}{(2\pi)^3}.
\end{equation}
Here we suppress the spin indices for simplicity. When the system is
near FCQPT, the approximate interpolative solution for Eq.
\eqref{sam2} reads \cite{epl2,prlmy,obz}
\begin{equation}
\frac{M^*(T_N,x)}{M^*_M}={M^*_N(T_N)}\approx
c_0\frac{1+c_1T_N^2}{1+c_2T_N^{8/3}}. \label{fin1}
\end{equation}
Here $M^*_N(T_N)$ is the normalized effective mass, $M^*_M$ is the
maximum value, that it reaches at $T=T_M$. Normalized temperature
$T_N=T/T_M$, $c_0=(1+c_2)/(1+c_1)$, $c_1$ and $c_2$ are fitting
parameters, parameterizing Landau amplitude. It follows from Eq.
\eqref{fin1} that in contrast to the standard paradigm of
quasiparticles the effective mass strongly depends on temperature,
revealing three different regimes at growing temperature. At the
lowest temperatures we have the LFL regime. Then the system enters
the transition regime: $M^*_N(T_N)$ grows, reaching its maximum
$M^*_N=1$ at $T=T_M$, ($T_N=1$), with subsequent diminishing. Near
temperatures $T_N\geq 1$ the last "traces" of LFL regime disappear
and the NFL state takes place, manifesting itself in decreasing of
$M^*_N$ as $T_N^{-2/3}$ and then as
\begin{equation}\label{r2}
    M^*_N(T_N) \propto \frac{1}{\sqrt{T_N}}.
\end{equation} These regimes are
reported in the inset to Fig. \ref{PHD}.

\begin{figure}[!ht]
\begin{center}
\includegraphics [width=0.45\textwidth]{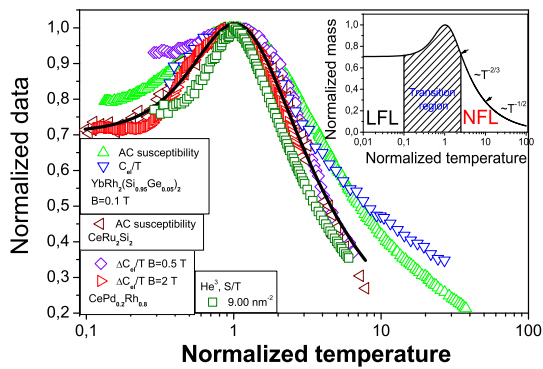}
\end{center}
\caption{The universal behavior of $M^*_N(T_N)$, extracted from
measurements of different thermodynamic quantities, as shown in the
legend. The $AC$ susceptibility, $\chi_{AC}(T,B)$, is taken for
YbRh$_2$(Si$_{0.95}$Ge$_{0.05}$)$_2$ and CeRu$_2$Si$_2$
\cite{geg3,takah}, the heat capacity divided by temperature, $C/T$,
is taken for YbRh$_2$(Si$_{0.95}$Ge$_{0.05}$)$_2$ and
CePd$_{0.2}$Rh$_{0.8}$ \cite{cust,pikul} and entropy divided by
temperature, $S/T$, for 2D $\rm ^3He$ is taken from Ref. \cite{he3}.
The solid curve gives the theoretical universal behavior of $M^*_N$
determined by Eq. \eqref{fin1}. Inset shows normalized effective
mass $M^*_N(T_N)$ \eqref{fin1} versus the normalized temperature
$T_N=T/T_M$. The hatched area outlines the transition regime.
Several regions are shown as explained in the text.}\label{FD}
\end{figure}

As it follows from Eq. \eqref{fin1}, $M^*$ reaches the maximum
$M^*_M$ at some temperature $T_M$. Since there is no external
physical scales near FCQPT point, the normalization of both $M^*$
and $T$ by internal parameters $M^*_M$ and $T_M$ immediately reveals
the common physical nature of above thermodynamic functions which we
use to extract the effective mass. The normalized effective mass
extracted from measurements on the HF metals
YbRh$_2$(Si$_{0.95}$Ge$_{0.05}$)$_2$ \cite{geg3,cust},
CeRu$_2$Si$_2$ \cite{takah}, CePd$_{1-x}$Rh$_x$ \cite{pikul}, and 2D
$\rm ^3He$ \cite{he3} along with our theoretical solid curve (also
shown in the inset) is reported in Fig. \ref{FD}. It is seen that
above normalization of experimental data yields the merging of
multiple curves into single one, thus demonstrating a universal
scaling behavior \cite{epl2,prlmy,prb}. It is also seen that the
universal behavior of the effective mass given by our theoretical
curve agrees well with experimental data.

It is seen from Fig. \ref{FD} that at $T/T_M=T_N\leq 1$ the
$T$-dependence of the effective mass is weak. This means that the
$T_M$ point can be regarded as a crossover between LFL and NFL
regimes. Since magnetic field enters the Landau equation as
$\mu_BB/T$, we have
\begin{equation}
T^*(B)=a_1+a_2B\simeq T_{M}\sim \mu_B
(B-B_{c0})\label{TB},
\end{equation}
where $T^*(B)$ is the crossover temperature, $\mu_B$ is Bohr
magneton, $a_1$ and $a_2$ are constants. The crossover temperature
is not really a phase transition. It necessarily is broad, very much
depending on the criteria for determination of the point of such a
crossover, as it is seen from the inset to Fig. \ref{FD}. As
usually, the temperature $T^*(B)$ is extracted from the field
dependence of charge transport, for example from the resistivity
$\rho(T)=\rho_0+A(B)T^2$ with $\rho_0$ is a temperature independent
part and $A(B)$ is a LFL coefficient. The crossover takes place at
temperatures where the resistance starts to deviate from the LFL
$T^2$ behavior, see e.g. Ref. \cite{pnas}. We note that Eq.
\eqref{MB} is valid at $T<T^*(B)$. In that case, magnetic field
plays a role of the control parameter $\zeta$ at fixed $T_N$ as
shown in Fig. \ref{PHD} by the dash horizontal arrow.

\begin{figure}[!ht]
\begin{center}
%\vspace*{-7cm}
\includegraphics [width=0.45\textwidth]{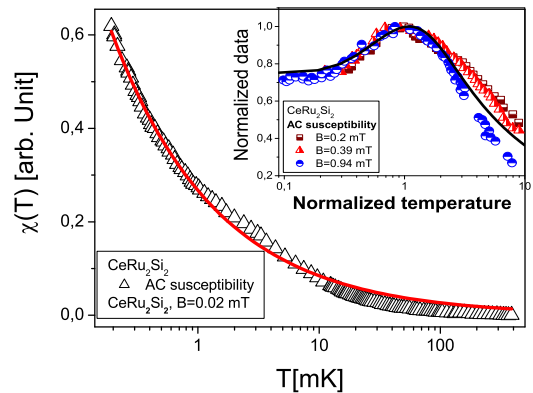}
\end{center}
\caption{Temperature dependence of the $AC$ susceptibility
$\chi_{AC}$ for $\rm{CeRu_2Si_2}$. The solid curve is a fit for the
data shown by the triangles at $B=0.02$ mT \cite{takah} and
represented by the function $\chi(T)=a/\sqrt{T}$ given by Eq.
\eqref{r2} with $a$ being a fitting parameter. Inset reports
$M^*_N(T_N)$ extracted from $\chi_{AC}$ measured at different fields
as indicated in the legend \cite{takah}. The solid curve traces the
universal behavior \eqref{fin1}. Parameters $c_1$ and $c_2$ are
adjusted to fit the average behavior of the normalized effective
mass $M^*_N$.}\label{MRM}
\end{figure}

To verify Eq. \eqref{r2} and illustrate the transition from LFL to
NFL regime, we use measurements of $\chi_{AC}(T)$ in
$\rm{CeRu_2Si_2}$ at magnetic field $B=0.02$ mT at which this HF
metal demonstrates the NFL behavior down to lowest temperatures
\cite{takah}. Indeed, in this case we expect that LFL regime emerges
at temperatures lower than $T_M\sim \mu_B B\sim 0.01$ mK as it
follows from Eq. \eqref{TB}. It is seen from Fig. \ref{MRM} that Eq.
\eqref{r2} gives good description of the facts in the extremely wide
temperature range: the susceptibility $\chi_{AC}$ as a function of
$T$, is not a constant upon cooling, as would be for a Fermi liquid,
but shows a $1/\sqrt{T}$ divergence over more than three decades in
temperature. The inset to Fig. \ref{MRM} exhibits a fit for $M^*_N$
extracted from measurements of $\chi_{AC}(T)$ at different magnetic
fields, clearly indicating the transition from LFL behavior at
$T_N<1$ to NFL one at $T_N>1$ when the system moves along the
vertical arrow in Fig. \ref{PHD}.  It seen from  Fig. \ref{MRM} that
the function given by Eq. \eqref{fin1} represents a good
approximation for $M^*_N$.

\section{Common field-induced quantum critical point in HTSC
compounds and HF metals}

\begin{figure}[!ht]
\begin{center}
%\vspace*{-0.5cm}
\includegraphics [width=0.45\textwidth]{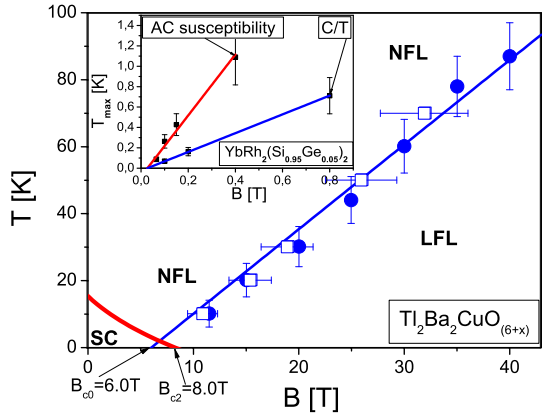}
\end{center}
\caption{$B-T$ phase diagram of superconductor
Tl$_2$Ba$_2$CuO$_{6+x}$. The crossover (from LFL to NFL regime)
line $T^*(B)$ is given by the Eq. \eqref{TB}. Open squares and
solid circles are experimental values \cite{pnas}. Thick line
represents the boundary between the superconducting and normal
phases. Arrows near the bottom left corner indicate the critical
magnetic field $B_{c2}$ destroying the superconductivity and the
critical field $B_{c0}$. Inset reports the peak temperatures
$T_{\rm max}(B)$, extracted from measurements of $C/T$ and
$\chi_{AC}$ on YbRh$_2$(Si$_{0.95}$Ge$_{0.05}$)$_2$
\cite{cust,geg3} and approximated by straight lines \eqref{TB}. The
lines intersect at $B\simeq 0.03$ T.}\label{TMM}
\end{figure}

Let us now consider the $B-T$ phase diagram of the HTSC substance
Tl$_2$Ba$_2$CuO$_{6+x}$ shown in Fig. \ref{TMM}. The substance is a
superconductor with $T_c$ from 15 K to 93 K depending on oxygen
content \cite{pnas}. In Fig. \ref{TMM}, open squares and solid
circles show the experimental values of the crossover temperature
from the LFL to NFL regimes \cite{pnas}. The solid line shows our
fit \eqref{TB} with $B_{c0}=6$ T that is in good agreement with
$B_{c0}=5.8$ T obtained from the field dependence of the charge
transport \cite{pnas}.
\begin{figure} [! ht]
\begin{center}
\vspace*{-0.5cm}
\includegraphics [width=0.44\textwidth]{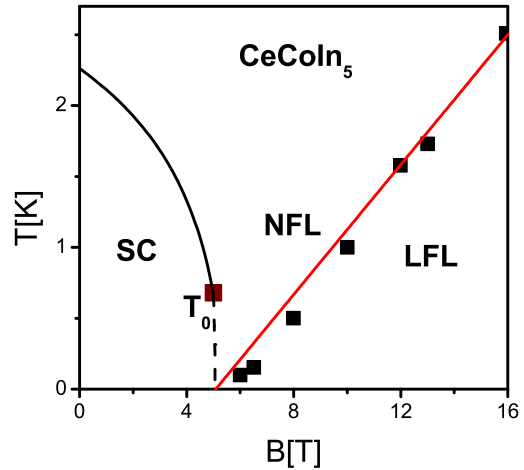}
\end{center}
\vspace*{-0.7cm} \caption{$B-T$ phase diagram of the $\rm CeCoIn_5$
heavy fermion metal. The crossover line between superconducting and
normal phases is shown by the solid line at $T>T_0$ and dashed one
at $T<T_0$. The point $T_0$, shown by square signifies a
temperature, where the phase transition becomes a first-order (at
$T<T_0$ \cite{bian}). The solid straight line specified by Eq.
\eqref{TB} with the experimental points \cite{pag1} shown by squares
is a boundary between Landau Fermi liquid (LFL) and non-Fermi-liquid
(NFL) states.}\label{CeCo}
\end{figure}
As it is seen from Fig. \ref{TMM}, the linear behavior agrees well
with experimental data \cite{pnas}. The peak temperatures $T_{\rm
max}$ shown in the inset to Fig. \ref{TMM}, report the maxima of
$C(T)/T$ and $\chi_{AC}(T)$ measured on
YbRh$_2$(Si$_{0.95}$Ge$_{0.05}$)$_2$ \cite{geg3,cust}. As it follows
from Eq. \eqref{TB}, $T_{\rm max}$ shifts to higher values with
increase of the applied magnetic field. It is seen that both
functions can be represented by straight lines intersecting at
$B\simeq 0.03$ T. This observation is in good agreement with
experiments \cite{geg3,cust}. It is seen from Fig. \ref{TMM} that
critical field $B_{c2}=8$ T destroying the superconductivity is
close to $B_{c0}=6$ T. Let us show that this is more than a simple
coincidence, and $B_{c2}\gtrsim B_{c0}$. Indeed, at $B>B_{c0}$ and
low temperatures $T<T^*(B)$, the system is in its LFL state. The
superconductivity is then destroyed since the superconducting gap is
exponentially small as we have seen above. At the same time, there
is FC state at $B<B_{c0}$ and this low-field phase has large
prerequisites towards superconductivity as in this case the gap is a
linear function of the coupling constant as it was also shown above.
We note that this is exactly the case in $\rm CeCoIn_5$ where
$B_{c0}\simeq B_{c2}\simeq 5$ T \cite{pag} as seen from Fig.
\ref{CeCo}, while the application of pressure makes $B_{c2}>B_{c0}$
\cite{ron}. On the other hand, if the superconducting coupling
constant is rather weak then antiferromagnetic order wins a
competition. As a result, $B_{c2}=0$, while $B_{c0}$ can be finite
as in $\rm YbRh_2Si_2$ and $\rm{YbRh_2(Si_{0.95}Ge_{0.05})_2}$
\cite{geg3,geg}.

Upon comparing the phase diagrams of Tl$_2$Ba$_2$CuO$_{6+x}$  and
$\rm CeCoIn_5$ (Figs. \ref{TMM} and \ref{CeCo} respectively), it is
possible to conclude that they are similar in many respects.
Further, we note that the superconducting boundary line $B_{c2}(T)$
at lowering temperatures acquires a step, i.e. the corresponding
phase transition becomes first order \cite{bian,epl}. This permits
us to speculate that the same may be true for
Tl$_2$Ba$_2$CuO$_{6+x}$. We expect that in the NFL state the
tunneling conductivity is asymmetric function of the applied
voltage, while it becomes symmetric at the application of elevated
magnetic fields when Tl$_2$Ba$_2$CuO$_{6+x}$ transits to the LFL
regime, as it predicted to be in $\rm CeCoIn_5$ \cite{pla}.

Now we consider the field-induced reentrance of LFL behavior in
Tl$_2$Ba$_2$CuO$_{6+x}$ at $B\geq B_{c2}$. In that case, the
effective mass $M^*$ depends on magnetic field $B$ taking the role
of the control parameter $\zeta$, while the system is in the LFL
regime as it is shown by the dashed horizontal arrow in Fig.
\ref{PHD}. The LFL regime is characterized by the temperature
dependence of the resistivity, $\rho(T)=\rho_0+A(B)T^2$, see also
above. The $A$ coefficient, being proportional to the
quasiparticle–--quasiparticle scattering cross-section, is found to
be $A\propto (M^*(B))^2$ \cite{geg}. With respect to Eq. \eqref{MB},
this implies that
\begin{equation}
A(B)\simeq
A_0+\frac{D}{B-B_{c0}},\label{HFTC}
\end{equation}
where $A_0$ and $D$ are parameters. It is pertinent to note that
Kadowaki-Woods ratio \cite{kw},  $K=A/\gamma_0^2$, is constant
within our FCQPT theory as it follows from Eqs. \eqref{MB} and
\eqref{HFTC} \cite{khz}. It follows from Eq. \eqref{HFTC} that it
is impossible to observe the relatively high values of $A(B)$ since
in our case $B_{c2}>B_{c0}$. We note that Eq. \eqref{HFTC} is
applicable when the superconductivity is destroyed by the
application of magnetic field, otherwise the effective mass is also
finite being given by Eq. \eqref{17}. Therefore, as was mentioned
above, in HTSC, a QCP is poorly accessible to experimental
observations being {\it "hidden in superconductivity"}.
Nonetheless, thanks to recent experimental facts \cite{pnas}, we
will see that it is possible to study QCP by exploring its
"shadows".

\begin{figure}[!ht]
\begin{center}
%\vspace*{-0.5cm}
\includegraphics [width=0.45\textwidth]{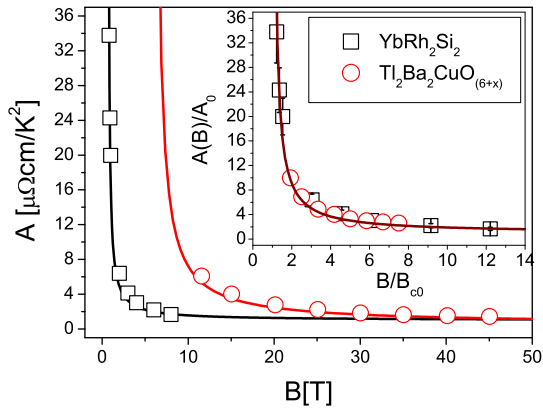}
\end{center}
\caption{The charge transport coefficient $A(B)$ as a function of
magnetic field $B$ obtained in measurements on YbRh$_2$Si$_2$ (
squares) \cite{geg} and Tl$_2$Ba$_2$CuO$_{6+x}$ (circles)
\cite{pnas}. The different field scales are clearly seen. In the
inset, normalized coefficient $A(B)/A_0\simeq 1+D_N/(y-1)$ as a
function of normalized magnetic field $y=B/B_{c0}$ is shown by
squares for YbRh$_2$Si$_2$ and by circles for
Tl$_2$Ba$_2$CuO$_{6+x}$. $D_N$ is the only fitting
parameter.}\label{f2}
\end{figure}

Figure \ref{f2} reports the fit of our theoretical dependence
\eqref{HFTC} to the experimental data for two different classes of
substances: HF metal YbRh$_2$Si$_2$ and HTSC
Tl$_2$Ba$_2$CuO$_{6+x}$. The different scale of fields is clearly
seen as well as good coincidence with theoretical dependence
\eqref{HFTC}. This means that the physics underlying the
field-induced reentrance of LFL behavior, is the same for both
classes of substances. To further corroborate this point, we rewrite
Eq. \eqref{HFTC} in reduced variables $A/A_0$ and $B/B_{c0}$. Such
rewriting immediately reveals the universal nature of the behavior
of these two substances - both of them are driven to common QCP
related to FC and induced by the application of magnetic field. As a
result,  Eq. \eqref{HFTC}  takes the form
\begin{equation} \frac{A(B)}{A_0}\simeq
1+\frac{D_N}{B/B_{c0}-1},\label{HFC}
\end{equation}
where $D_N=D/(A_0B_{c0})$ is a constant. It is seen from Eq.
\eqref{HFC} that upon applying the scaling, the quantities $A(B)$
for $\rm Tl_2Ba_2CuO_{6+x}$ and $\rm YbRh_2Si_2$ are reduced to a
function of the single variable $B/B_{c0}$ thus demonstrating
universal behavior. To support Eq. \eqref{HFC}, we replot both
dependencies in reduced variables $A/A_0$ and $B/B_{c0}$ as it is
depicted in the inset to Fig. \ref{f2}. Such replotting immediately
reveals the universal nature of the behavior of these two
substances. It is seen from the inset to Fig. \ref{f2} that close to
magnetic QCP there is no external physical scales so that the
normalization by internal scales $A_0$ and $B_{c0}$ shows
straightforwardly the common physical nature of these substances
behavior.

\section{Summary}

Our comprehensive theoretical study of vast majority of experimental
facts regarding very different strongly correlated Fermi-systems
such as high-temperature superconductors, heavy-fermion compounds
and two-dimensional $\rm ^3He$ clearly demonstrates their generic
family resemblance. We show that the physics underlying the
field-induced reentrance of LFL behavior is the same for both HTSC
compounds and HF metals. We also show that there is a relationship
between the critical fields $B_{c2}$ and  $B_{c0}$ so that
$B_{c2}\gtrsim B_{c0}$. It follows from our study that there is at
least one quantum phase transition inside the superconducting dome,
and this transition is the fermion condensation quantum phase
transition.

This work was supported in part by the grants: RFBR No. 09-02-00056.

\end{document}